\documentclass[aps,prl,reprint,superscriptaddress]{revtex4-2}

\usepackage{graphicx}
\usepackage{amsmath}
\usepackage{soul}
\usepackage{color}
\usepackage{hyperref}

\begin{document}


\title{Structural signatures of primary and secondary relaxation during physical aging of a metallic glass}


\author{Beatrice Baraldi}
\author{Stefano Marchesin}
\author{Davide Marchesi}
\author{Lara Piemontese}
\author{Francesco Dallari}
\affiliation{Dipartimento di Fisica e Astronomia ``Galileo Galilei'', Universit\`a degli Studi di Padova}

\author{S{\o}ren S.\ S{\o}rensen}
\affiliation{Department of Chemistry and Bioscience, Aalborg University, Aalborg, Denmark}

\author{Eloi Pineda}
\affiliation{Department of Physics, Institute of Energy Technologies, Barcelona Research Centre in Multiscale Science and Engineering, Universitat Polit{\`e}cnica de Catalunya—BarcelonaTech, Barcelona, Spain}

\author{Marco Di Michiel}
\affiliation{ESRF -- The European Synchrotron, 71 Avenue des Martyrs, CS40220, 38043 Grenoble Cedex 9, France}

\author{Giulio Monaco}
\author{Peihao Sun}
\email{peihao.sun@unipd.it}
\affiliation{Dipartimento di Fisica e Astronomia ``Galileo Galilei'', Universit\`a degli Studi di Padova}

\date{\today}

\begin{abstract}
Physical aging drives glasses toward equilibrium and changes their thermodynamic and mechanical properties, but the accompanying microscopic structural evolution remains poorly understood. Here, using our recently developed low-background setup combining fast scanning calorimetry with \emph{in-situ} X-ray total scattering, we investigate microscopic structural changes during physical aging in Pd$_{42.5}$Cu$_{30}$Ni$_{7.5}$P$_{20}$ metallic glass. The measurements span an unprecedented six orders of magnitude in time and a temperature range extending from above the glass transition temperature to more than 150\,K below it. We observe dominant structural changes near the first and second peaks of the structure factor. Near and above the glass transition temperature, changes in the first sharp diffraction peak closely follow enthalpy relaxation and are associated predominantly with increasing degrees of medium-range ordering. At lower temperatures, the enthalpy evolution can no longer be described by this contribution alone, revealing an additional, faster relaxation process that correlates with changes near the second diffraction peak and thus short-range structural rearrangements, including increased Pd--Pd correlations. These results establish a direct connection between thermodynamic relaxation and structural evolution, providing strong evidence that primary and secondary aging processes are associated with distinct structural length scales.

\end{abstract}

\maketitle

\section{Introduction}
Aging is a universal phenomenon in glasses where, after the system falls out of equilibrium upon cooling below the glass transition temperature, it continues to evolve with time toward equilibrium~\cite{Cangialosi2024PhysicalEffects,Micoulaut2016RelaxationReview,Ruta2017RelaxationGlasses,Gallino2024PhysicalGlasses}. This evolution is accompanied by changes in various macroscopic properties such as density, enthalpy, and viscosity. In metallic glasses, physical aging can also alter mechanical properties, including the elastic moduli~\cite{Wang2012TheGlasses} and brittleness~\cite{Lewandowski2005IntrinsicGlasses}, and is therefore an important factor to consider for technological and industrial applications. Despite extensive research on this topic, many aspects of physical aging remain unclear because of its intrinsic complexity and non-equilibrium nature~\cite{Cangialosi2024PhysicalEffects}. In particular, the microscopic structural changes during aging are not yet fully understood, reflecting the broader and long-standing quest to establish structure--property relations in glassy systems. Several previous studies~\cite{Mao1995X-ray-diffractionGlasses,Mattern2012ThermalGlass,Giordano2015} have examined the effects of annealing, aging, and cooling rate on the microscopic structure of metallic glasses, but a coherent picture has yet to emerge, and the connection between structural evolution and changes in macroscopic properties such as enthalpy remains unclear.

In this work, we combine X-ray total scattering with fast scanning calorimetry (FSC) to perform \emph{in-situ} structural measurements during physical aging over a broad range of temperatures and time scales. The use of FSC provides rapid and precise temperature control, enabling reproducible measurements on time scales as short as milliseconds. In comparison, conventional temperature control systems usually operate at heating and cooling rates on the order of 20\,K/min, thereby setting a lower limit for the measurable time to be minutes or longer. Hence, the use of FSC provides several orders of magnitude of improvement in the range of probed time scales, making it possible to study aging-induced relaxation processes in detail, as will be shown below. Moreover, our experimental setup allows low-background X-ray diffraction measurements at momentum transfers ($Q$) from below 1\,\r{A}$^{-1}$ to above 18\,\r{A}$^{-1}$~\cite{Sun2025ACalorimetry}. This makes it possible to reliably and reproducibly obtain the structure factor and, subsequently, the pair distribution function (PDF) of the sample. Therefore, combining these two techniques, our setup is able to provide direct and quantitative insight into the microscopic structural evolution that occurs during aging with an unprecedented range of time scales. 

As a model system, we investigate the metallic glass with composition Pd$_{42.5}$Cu$_{30}$Ni$_{7.5}$P$_{20}$ (at.\%). This alloy has been extensively characterized in the literature and is found to exhibit high glass-forming ability together with several notable properties, including a pronounced secondary ($\beta$) relaxation below the glass transition temperature, $T_g$~\cite{Qiao2014CharacteristicsLiquids}. After cooling the sample from the equilibrium supercooled liquid state well above $T_g$, we perform X-ray diffraction measurements as a function of time while keeping the sample at a constant aging temperature. The accessible aging times extend from tens of milliseconds to hours, spanning almost six orders of magnitude, while the investigated aging temperatures range from more than 30\,K above $T_g$ to more than 150\,K below it. These measurements provide a comprehensive picture of microscopic structural evolution during physical aging and reveal signatures of different relaxation processes at distinct structural length scales.

\section{Results}
\subsection{Changes in the structure factor}
\begin{figure*}
    \centering
    \includegraphics{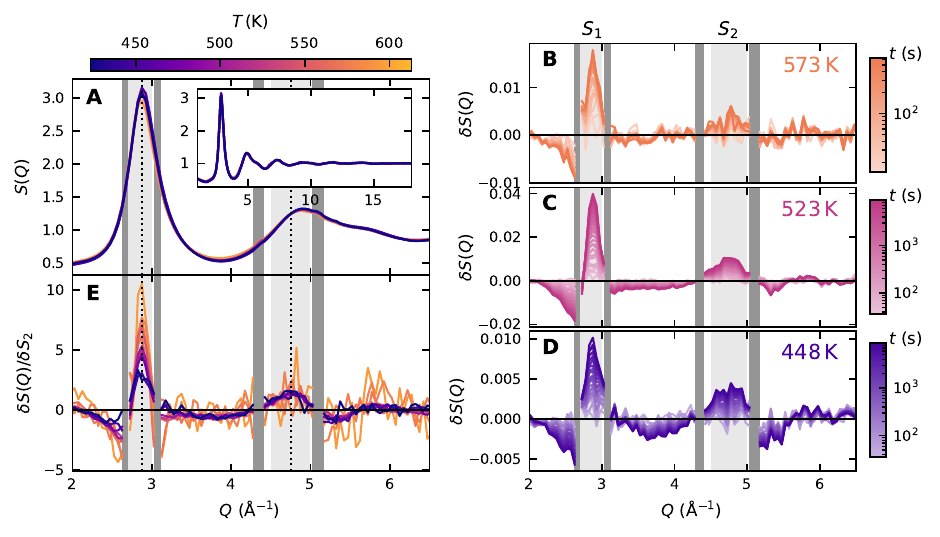}
    \caption{\textbf{Evolution of the structure factor during aging.} 
    (A) Time-averaged structure factor, $S(Q)$, at various temperatures indicated by the color bar. The inset shows the full measured $Q$-range. Dark-gray shaded regions mark $Q$-ranges potentially influenced by Bragg reflections from gold from the FSC chip and therefore excluded from the analysis. 
    (B--D) Aging-induced changes in $S(Q)$ at three representative temperatures. The curves show the evolution relative to the initial measurement at each temperature. Light-gray shaded regions identify the ranges used to calculate the changes associated with the first- and second-peak regions, $\delta S_1$ and $\delta S_2$, respectively. 
    (E) Aging-induced changes in the structure factor, $\delta S(Q)$, normalized by the integrated change in the second-peak region, $\delta S_2$. The relative contribution of the first-peak region increases with increasing temperature.}
    \label{fig:SQ}
\end{figure*}

We first examine the evolution of the structure factor, $S(Q)$, during aging at different temperatures. Figure~\ref{fig:SQ}A shows the time-averaged $S(Q)$ at each temperature over the $Q$-range encompassing the first two diffraction peaks; the inset shows the full $Q$-range from 1 to 18\,\r{A}$^{-1}$. Except for small shifts in the peak positions, consistent with previous results~\cite{Georgarakis2011VariationsVitrification,Sun2025ACalorimetry}, the curves are similar.

However, differences emerge in the time evolution of the structure factor at different aging temperatures, $T$. Figure~\ref{fig:SQ}B--D shows the time evolution of $S(Q)$ during aging at three representative temperatures. Here we have excluded data corresponding to the $Q$-positions of the Debye-Scherrer rings of elemental gold present on the active area of the FSC chip (see Methods and Ref.~\cite{Sun2025ACalorimetry} for further information), as shown by the dark gray shaded regions in Figure~\ref{fig:SQ}A. Several features are notable. First, at each $T$, $S(Q)$ evolves gradually but steadily with time. The most prominent changes in $S(Q)$ are gradual increases in two $Q$-ranges around 2.87\,\r{A}$^{-1}$ and 4.75\,\r{A}$^{-1}$, as indicated by the light gray shaded regions denoted as $S_1$ and $S_2$ henceforth. Across different temperatures, the overall shape and positions are similar, but at lower $T$, changes in the second peak $S_2$ become more significant compared to those in the first peak $S_1$.

To more clearly compare the structural evolution at different temperatures, we normalize the change $\delta S(Q)$ by the integrated change in the second peak, $\delta S_2 = \int_{Q_\mathrm{2min}}^{Q_\mathrm{2max}} \Delta S(Q) dQ/(Q_\mathrm{2max}-Q_\mathrm{2min})$, where $Q_\mathrm{2min,2max}=4.5,5.0$\,\r{A}$^{-1}$. The results, shown in Figure~\ref{fig:SQ}E, demonstrate a clear trend: changes in the first peak become more pronounced at higher $T$. Although the choice of start and end times for the calculation of the difference $\Delta S(Q)$ is somewhat arbitrary, we note that, to first order, the changes appear to be single-mode at each $T$; for example, in each panel of Figure~\ref{fig:SQ}B--D, the change in $S_1$ appears to be roughly proportional to that in $S_2$ (see Figure~\ref{fig:slope_dS2dS1}). Therefore, the choice of start and end times does not influence the result qualitatively.

Notably, the change $\delta S(Q)$ has a shape distinct from that of $S(Q)$ itself. In particular, the second peak of $\delta S(Q)$ is centered at a $Q$-value significantly lower than that of $S(Q)$, while the first peaks are at similar positions, as indicated by the dotted vertical lines in Figure~\ref{fig:SQ}.
Since it is known that the first sharp peak in $S(Q)$ is related to medium-range ordering while higher-order peaks reflect ordering on a shorter range~\cite{Cargill1975}, our observations suggest that aging at lower temperatures is increasingly characterized by short-range ordering. Further details and implications will be discussed below.

Having identified $S_1$ and $S_2$ as the most prominent structural changes during aging, we now focus on quantitative analysis of these two regions as a function of time and temperature. Similarly to $\delta S_2$, we define the quantity $\delta S_1 = \int_{Q_\mathrm{1min}}^{Q_\mathrm{1max}} \delta S(Q) dQ/(Q_\mathrm{1max}-Q_\mathrm{1min})$, where $Q_\mathrm{1min,1max}=2.7,3.0$\,\r{A}$^{-1}$. As shown in Figure~\ref{fig:S1_combined} and Figure~\ref{fig:S2_combined}, the quantities $\delta S_1$ and $\delta S_2$ thus obtained show high reproducibility among samples on different FSC chips and scans performed during different experiments. We therefore combine all datasets for each aging temperature, covering almost six orders of magnitude in time scales.

\subsection{Primary relaxation and changes in the first peak}
\begin{figure}
    \centering
    \includegraphics{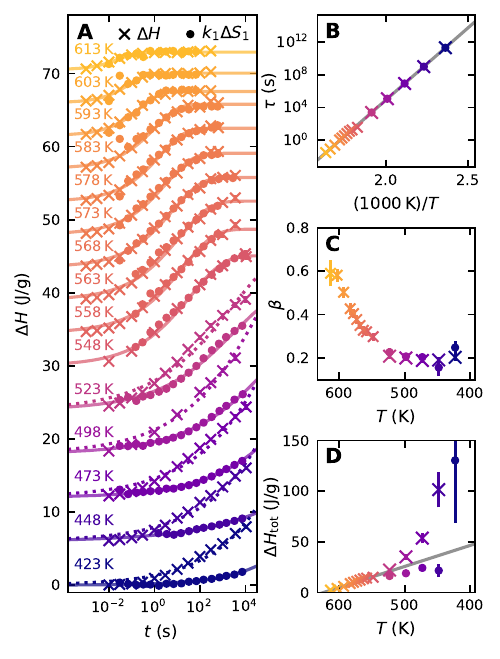}
    \caption{\textbf{Relationship between enthalpy relaxation and changes in the first-peak region.}
    (A) Enthalpy change, $\Delta H$, measured at different aging temperatures (crosses, $\times$), and the scaled structural change in the first-peak region, $k_1\Delta S_1$ (filled circles, $\bullet$). Solid and dotted curves show KWW fits.
    (B--D) Temperature dependence of the fitted KWW relaxation time, $\tau$, stretching exponent, $\beta$, and total relaxation amplitude, $\Delta H_{\mathrm{tot}}$, respectively. Crosses represent parameters obtained from fits to $\Delta H$, whereas filled circles represent parameters obtained from fits to $k_1\Delta S_1$ at $T < 548$\,K. The solid line in panel (B) is an Arrhenius fit to the relaxation times. The solid line in panel (D) is a linear fit to the amplitudes measured at $T \geq 548$\,K.}
    \label{fig:S1_vs_H}
\end{figure}

We now examine structural changes throughout the aging process at different temperatures. To characterize the state of aging, we measure the change in enthalpy, $\Delta H$, as a function of time with FSC. $\Delta H(t)$ is defined as the integral of the difference in heat capacity without aging and after aging for a period of time $t$. The results are plotted as crosses in Figure~\ref{fig:S1_vs_H}. We then fit the data at each temperature $T$ with a Kohlrausch--Williams--Watts (KWW) function, as has been done in previous works~\cite{Gallino2024PhysicalGlasses}:
\begin{equation} \label{eq:KWW}
    \Delta H(t) = \Delta H_\text{tot}(T) \left[1 - e^{-(t/\tau)^\beta} \right],
\end{equation}
where $\Delta H_\text{tot}$ is the total amplitude of the change, $\tau$ is the characteristic time, and $\beta$ is the stretching exponent ($0 < \beta < 1$).

We first focus on the higher aging temperatures, $T \geq 548$\,K. As shown in Figure~\ref{fig:S1_vs_H}A, the KWW fits describe well the $\Delta H (t)$ data in this temperature range. The extracted KWW parameters, shown as crosses in panels B-D, exhibit the expected temperature dependence: with decreasing $T$, the characteristic time $\tau$ becomes longer, the amplitude $\Delta H_\text{tot}$ increases, and the relaxation becomes more stretched (smaller $\beta$)~\cite{Gallino2024PhysicalGlasses}. The relaxation time appears to follow the Arrhenius equation: $\tau = \tau_0 \exp(E_a/R T)$, where $\tau_0$ is a constant, $E_a$ the activation energy, and $R$ the gas constant; from a fit to the data (shown by the line in Figure~\ref{fig:S1_vs_H}B), we can extract an activation energy of $(336 \pm 10)$\,kJ/mol. The amplitude $\Delta H_\text{tot}$ instead appears to be linear with temperature, and a linear fit to the data yields $\Delta H_\text{tot}(T) = (624\,\mathrm{K} - T) \cdot (0.21\,\mathrm{J\,g^{-1}\,K^{-1}}$). This suggests that the system begins to fall out of equilibrium around 624\,K with the selected cooling rate of $10^4$\,K/s.
The slope, on the other hand, agrees with the heat capacity difference between the supercooled liquid and the glassy states~\cite{Haruyama2014VolumeGlass}. These results confirm the commonly held picture for physical aging where the system eventually reaches the equilibrium supercooled liquid state that is the linear extension of the states above $T_f$~\cite{Cangialosi2024PhysicalEffects}.

The structural data for higher temperatures ($T \geq 548$\,K) closely follow the change in enthalpy. To demonstrate this, we fit the enthalpy and structural data in this temperature range with a simple linear model:
\begin{equation} \label{eq:model_S1}
    \Delta H(T, t) = k_1 \Delta S_1(T, t),
\end{equation}
where $k_1$ is a universal coefficient for all $T \geq 548$\,K. The change in $S_1$ is defined as $\Delta S_1(T, t) \equiv \delta S_1(T, t) - \delta S_1(T, t = 0)$, where $\delta S_1(T, t = 0)$ is left as a free parameter in the fit because its direct determination requires prohibitively high counting statistics within a very short time window; this amounts to a possible overall shift in the $S_1$ data at each temperature. We note that data at lower temperatures are not considered for fitting, for reasons that will become clear in the next section. The fit result for $k_1 \Delta S_1$ using this model is shown as filled circles in Figure~\ref{fig:S1_vs_H}A, and it agrees well with the enthalpy change. The conversion coefficient thus obtained is $k_1 = (294 \pm 5)$\,J/g. Therefore, at temperatures close to and above $T_g$, our data indicate a simple linear relation between the primary relaxation (represented here as enthalpy relaxation) and changes in the structure factor.

\subsection{Sub-$T_g$ aging and secondary relaxation}
Next, we turn to aging at lower temperatures, $T < 548$\,K, where the behavior changes in several aspects. First, the simple linear relation between enthalpy and structural changes proposed above no longer holds: as shown in Figure~\ref{fig:S1_vs_H}A, the difference between $\Delta H$ and $k_1 \Delta S_1$ keeps increasing as $T$ decreases, and it clearly cannot be attributed to a baseline shift in $\Delta S_1$. The data suggest that enthalpy relaxation happens faster than the intensification of the first sharp diffraction peak.

Second, the KWW model no longer provides a physically meaningful description of $\Delta H(t)$. Demonstrating this is not straightforward because complete relaxation at these low temperatures would require extremely long times far beyond the experimental window. Thus, only the initial part of the relaxation can be measured, which leaves too much freedom in the KWW parameters. For example, in the absence of constraints, the best fit yields a characteristic time ($\tau$) much shorter than that predicted by the Arrhenius model, which is physically implausible. On the other hand, if we constrain $\tau$ to take the value extrapolated from the Arrhenius relation at high temperatures, the resulting fit (shown as dotted lines in Figure~\ref{fig:S1_vs_H}) does not describe the data well; more importantly, the fitted amplitude $\Delta H_\text{tot}$ increases exponentially with decreasing $T$ (see Figure~\ref{fig:S1_vs_H}D), which is also physically implausible. By contrast, the KWW model can still describe the changes in $S_1$: using the coefficient $k_1$ determined at higher temperatures, the $\Delta S_1$ data (closed circles in Figure~\ref{fig:S1_vs_H}A) can be fitted with a KWW function (solid lines in Figure~\ref{fig:S1_vs_H}A), with $\tau$ fixed by the Arrhenius relation. The resulting values of $\Delta H_\text{tot}$ lie somewhat below the linear extrapolation from higher temperatures (see Figure~\ref{fig:S1_vs_H}D), a result that is much more plausible physically and will be discussed further below. The only exception is at the lowest temperature $T=423$\,K, where the data represents only a small part of the relaxation process and the fitted parameters consequently have large uncertainties.

Third, for $T < 548$\,K, changes in the second peak region $S_2$ become increasingly prominent relative to those in the first peak region $S_1$ (see Figure~\ref{fig:slope_dS2dS1}B). Note that since $S_1$ and $S_2$ are not completely correlated, the linear fit serves only as an illustration of the temperature-dependent trend.

\begin{figure}
    \centering
    \includegraphics{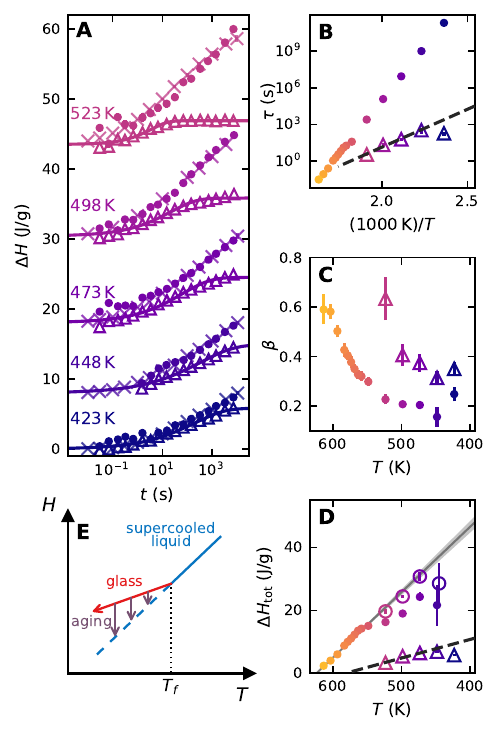}
    \caption{\textbf{Secondary relaxation revealed by the residual enthalpy change at low temperatures.}
    (A) Enthalpy change, $\Delta H$, during low-temperature aging (crosses, $\times$), residual enthalpy change after subtraction of the first-peak contribution, $\Delta H-k_1\Delta S_1$ (triangles, $\triangle$), and predictions of the structural model in Eq.~(\ref{eq:model_S2}) (filled circles, $\bullet$). Solid curves are KWW fits to the residual enthalpy change. (B--D) Temperature dependence of the KWW relaxation time, $\tau$, stretching exponent, $\beta$, and total relaxation amplitude, $\Delta H_{\mathrm{tot}}$, respectively. Filled circles reproduce the parameters obtained from the first-peak contribution, corresponding to the filled circles at $T < 548$\,K and the crosses at $T \geq 548$\,K in Figure~\ref{fig:S1_vs_H}B--D. Triangles represent parameters obtained from fits to the residual enthalpy change in panel (A). The dashed line in panel (B) is an Arrhenius fit to the triangle data, excluding the point at 423\,K. The dashed line in panel (D) is a linear fit to the corresponding residual amplitudes, also excluding 423\,K. The solid line in panel (D) is the linear fit to the high-temperature amplitudes at $T \geq 548$\,K, and the shaded region indicates one standard error. Open circles ($\bigcirc$) show the sum of the primary- and secondary-relaxation amplitudes. For clarity, the circles and filled circles at 448\,K are slightly offset horizontally. (E) Schematic diagram showing the commonly accepted picture of the enthalpy evolution in a glass-forming system in various states.}
    \label{fig:secondary}
\end{figure}

Together, these observations suggest a second, faster relaxation process at lower temperatures: this process contributes to $\Delta H$, which captures all relaxation processes, but is largely absent from $S_1$ which we associate mainly with primary relaxation.

To more quantitatively analyze this second component, we subtract $k_1 \Delta S_1$ from $\Delta H$, and the residuals are shown as triangular symbols in Figure~\ref{fig:secondary}A. These residuals appear to be well described by stretched exponentials, as shown by the KWW fit (solid lines). The extracted KWW parameters also suggest a secondary relaxation process that is independent of the primary relaxation. Apart from the lowest temperature $T=423$\,K for which only the initial relaxation is within the experimental window, the KWW fit parameters suggest another Arrhenius process (dashed line in Figure~\ref{fig:secondary}B) with an amplitude $\Delta H_\text{tot}$ that increases approximately linearly with as temperature decreases (dashed line in Figure~\ref{fig:secondary}D). A linear fit to $\Delta H_\text{tot}$ results in a vanishing amplitude around 582\,K, close to the glass transition temperature. Moreover, the activation energy extracted from the Arrhenius fit is $(119\pm 10)$\,kJ/mol, or around $(24.6 \pm 2.1) R T_g$, close to the commonly cited value of approximately $26 R T_g$ for the Johari-Goldstein $\beta$-relaxation process~\cite{Yu2013}. In addition, if we add the amplitude to that of the KWW fit to changes in $S_1$ obtained above, the sum, shown as empty circles in Figure~\ref{fig:secondary}D, is consistent with the linear extrapolation of the amplitude from high-temperature data where the primary relaxation dominates. This is in accordance with the generally held picture that the final state of physical aging is the equilibrium supercooled liquid state (see, e.g., Ref.~\cite{Cangialosi2024PhysicalEffects}), and thus the total enthalpy change during aging scales linearly with temperature below $T_f$, as shown by the schematic diagram in Figure~\ref{fig:secondary}E.

Regarding the structural origin of this secondary process, our observation that changes in the second peak $S_2$ begin to grow concomitantly with the gradual separation between $\Delta H$ and $\Delta S_1$ suggests that $S_2$ may be associated with the secondary process. We therefore expand Eq.~(\ref{eq:model_S1}) to include a contribution from $S_2$ as well:
\begin{equation} \label{eq:model_S2}
    \Delta H(T, t) = k_1 \Delta S_1(T, t) + k_2 [\Delta S_2(T, t)- c \Delta S_1(T, t)],
\end{equation}
where $k_2$ and $c$ are universal coefficients. The coefficient $c$ accounts for the part of $\Delta S_2$ attributed to primary relaxation. As shown in Figure~\ref{fig:secondary}A, this model can describe the data well, with extracted coefficients $k_2=(2070 \pm 120)$\,J/g and $c=0.256 \pm 0.010$. As expected, the value of $c$ is close to the ratio $\Delta S_2/\Delta S_1$ at high temperatures (see Figure~\ref{fig:slope_dS2dS1}B) where the primary relaxation dominates.

It is worth noting that the present results do not exclude the existence of secondary relaxation above 548\,K. However, the results suggest that its characteristic time is short and its relaxation amplitude small at 548\,K and above: the Arrhenius fit and the linear fit for the secondary process (shown as dashed black lines in Figure~\ref{fig:secondary}B and D) give $\tau = (1.1 \pm 0.4)$\,s and $\Delta H_\mathrm{tot} = (2.0 \pm 0.5)$\,J/g at 548\,K, respectively. Therefore, it is difficult to distinguish the secondary process from a constant background at these temperatures.

\section{Discussion}
\begin{figure}
    \centering
    \includegraphics{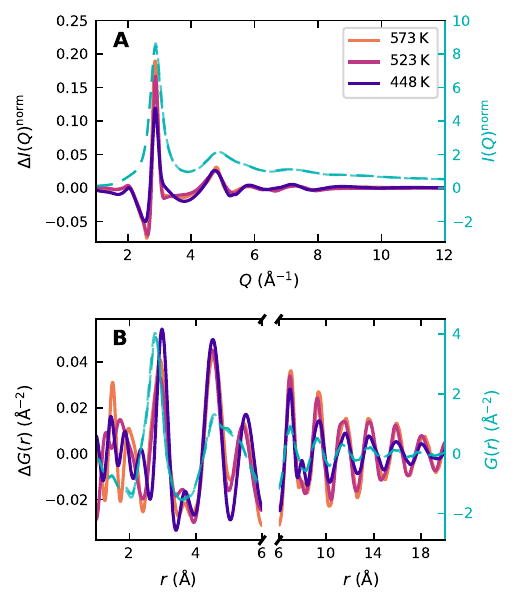}
    \caption{\textbf{Real-space signatures of the aging-induced structural changes.} (A) Normalized intensity profile, $I(Q)$, for the average structure at three representative aging temperatures (dashed curves, right axis), together with the corresponding normalized aging-induced change, $\Delta I(Q)$ (solid curves, left axis). (B) Pair distribution function, $G(r)$, for the average structure at the same temperatures (dashed curves, right axis), together with the corresponding aging-induced change, $\Delta G(r)$ (solid curves, left axis). The changes are normalized using the intensity variation in the $Q$-range associated with $S_2$, as described in the Methods section.}
    \label{fig:PDF}
\end{figure}

The results above identify distinct structural signatures associated with primary and secondary relaxations: changes near $S_1$ dominate the primary process, while at temperatures significantly below $T_g$, a secondary process emerges and is found to correlate with changes near $S_2$. The interpretation of the structural changes reflected in $S_1$ and $S_2$ merits further discussion. To help understand the associated structural changes in real space, we extract the change in the intensity profile, $\Delta I(Q)$, and the related change in the pair distribution function (PDF), $\Delta G(r)$, where $r$ is the pair distance, as described in the Methods section. Because in the regime of small variations, $\Delta G(r)$ is proportional to $\Delta I(Q)$, we normalize $I(Q)$ so that its integral is 1\% that of the baseline $I(Q)$ in the $Q$-range corresponding to $S_2$. Figure~\ref{fig:PDF} shows the differences $\Delta I(Q)$ and $\Delta G(r)$ together with the baselines $I(Q)$ and $G(r)$ corresponding to the three representative temperatures in Figure~\ref{fig:SQ}B--D. Note that normalizing to $S_2$ means that $\Delta S_1$ is reduced at lower temperatures (cf.\ Figure~\ref{fig:SQ}E). 

Generally speaking, in amorphous metal--metalloid alloys, the first sharp peak in the diffraction pattern is known to be mainly associated with medium-range ordering (MRO) beyond the second coordination shell~\cite{Cargill1975}. Similar sharpening of the first peak in $S(Q)$ upon annealing has been observed in various metallic glasses~\cite{Mao1995X-ray-diffractionGlasses,Mattern2012ThermalGlass}. Other studies have also reported densification during primary relaxation, for example from dilatometric measurements~\cite{Evenson2011EquilibriumGlass} or from changes in the position of the first sharp diffraction peak~\cite{Giordano2015}, which we denote as $Q_{m1}$.
In our case, however, $Q_{m1}$ cannot be directly related to the specific volume. Similar to previous observations in Pd$_{40}$Cu$_{30}$Ni$_{10}$P$_{20}$~\cite{Mattern2012ThermalGlass}, aging leads to an apparent shift of the first sharp diffraction peak toward lower $Q$, on the order of $\Delta Q_{m1} / Q_{m1} \sim 10^{-4}$, which may arise from the asymmetry of the peak. By contrast, the intensity changes provide a clear structural signature: $\Delta S_1$ is centered at approximately the same position as the first peak in $S(Q)$ (see Figure~\ref{fig:SQ}), while Figure~\ref{fig:PDF}B shows a corresponding increase in $G(r)$ at higher-order coordination shells ($r>6$\,\r{A}), whose amplitude increases with temperature. Taken together, these observations suggest that primary relaxation is characterized predominantly by an increase in MRO, reflected by enhanced correlations in higher-order coordination shells and the intensification of $S_1$.

For $S_2$, on the other hand, the interpretation is less straightforward. For many metallic liquids and glasses, the second peak in $I(Q)$ and its shoulder on the high-$Q$ side have been attributed to icosahedral short-range ordering (SRO)~\cite{Sachdev1984}. However, we note that the change $\Delta S(Q)$ has a shape rather different from that of $S(Q)$ around the second peak, as shown in Figure~\ref{fig:SQ}. In particular, the center of $\Delta S_2$ is significantly lower than that of the second peak, and quantitatively close results have been reported on a Pd$_{40}$Cu$_{30}$Ni$_{10}$P$_{20}$ sample~\cite{Mattern2012ThermalGlass}. Figure~\ref{fig:PDF}B shows that this feature corresponds to an increase on the high-$r$ side of the first coordination shell in $G(r)$. The peak position of this increase shifts from about 2.90\,\r{A} at 448\,K to about 2.98\,\r{A} at 573\,K. According to a comprehensive structural study of the same alloy as in the present work, the only comparable bond length is that of Pd--Pd~\cite{Hosokawa2022RelationshipGlass}. Therefore, our results suggest an enhancement of Pd--Pd correlations in the SRO with aging, particularly associated with the secondary relaxation process discussed above. Since this is the longest bond on average among the atomic pairs present in the material~\cite{Hosokawa2022RelationshipGlass}, it leads to an apparent expansion of the first coordination shell. For the second to third coordination shell, we observe peaks in $G(r)$ around 4.5\,\r{A} and 5.4\,\r{A}, which may correspond to enhanced Pd--Cu correlations~\cite{Hosokawa2022RelationshipGlass}.

The characteristics of different relaxation processes and their influence on material properties have been a topic of intense discussion in amorphous systems in general~\cite{Ngai2023UniversalApplications} and in metallic glasses in particular~\cite{Wang2019}. Here, our results provide strong evidence for two distinct structural relaxation processes, reflected in $S_1$ and $S_2$, that are related to changes in MRO and SRO, respectively. It is natural to attribute them to the $\alpha$- and $\beta$-relaxations widely discussed in the literature~\cite{Wang2019}, especially considering the temperature dependence of the two processes discussed above. 
Even though the secondary relaxation time observed here appears 1--2 orders of magnitude longer than previously reported $\beta$-relaxation times~\cite{Yu2012CorrelationGlasses,Qiao2014CharacteristicsLiquids}, we note that previous studies typically used dynamic mechanical measurements to probe relaxation, which can result in different characteristic time scales. For example, the characteristic $\alpha$-relaxation times are found to be about two orders of magnitude higher in calorimetric measurements than in dynamic shear modulus measurements in a Pd$_{40}$Ni$_{40}$P$_{20}$ glass-forming alloy~\cite{Wilde2001DirectTransition}. Similarly, in a Zr$_{46.8}$Ti$_{8.2}$Cu$_{7.5}$Ni$_{10}$Be$_{27.5}$ glass-forming alloy, the $\alpha$-relaxation times obtained from calorimetric measurements or dynamical measurements via X-ray photon correlation spectroscopy are found to be 1--2 orders of magnitude longer than those obtained from dynamic mechanical measurements~\cite{Amini2021}. In our case, X-ray diffraction is more sensitive to atomic species with higher atomic numbers, while it was found that the $\beta$-relaxation measured mechanically tended to correlate with the self-diffusion of the smallest constituent atom~\cite{Yu2012CorrelationGlasses} which is likely to have much faster dynamics~\cite{Amini2021,Gallino2018HierarchicalFormer}. Therefore, it is plausible that the secondary relaxation measured in the present work corresponds to the same process as the $\beta$-relaxation observed in mechanical measurements, but reflects different aspects of SRO during relaxation. 

It is worth noting that, although it is generally accepted that $\alpha$-relaxation involves many more atoms than $\beta$-relaxation, estimates of the corresponding structural length scales vary by several orders of magnitude among studies~\cite{Wang2019}. A recent work~\cite{Yang2022StructuralGlass} on a La-based metallic glass has found a link between $\beta$-relaxation and SRO. Therefore, the results in the present study provide direct structural evidence of the length scales involved in primary and secondary relaxations through aging.

In conclusion, the present results show that physical aging involves more than a single structural relaxation pathway. The primary process is associated predominantly with the growth of MRO, whereas the additional low-temperature contribution is linked to structural changes on shorter length scales corresponding to SRO. These two contributions are found to exhibit distinct structural signatures, characteristic time scales, and temperature dependences. Taken together, these observations provide a microscopic structural basis for the understanding of primary and secondary relaxation during aging. More broadly, they indicate that the hierarchy of relaxation dynamics in metallic glasses is accompanied by a corresponding hierarchy of structural rearrangements.

\begin{acknowledgments}
\paragraph{Acknowledgements} This project was funded by the European Union -- NextGenerationEU, under the Italian National Recovery and Resilience Plan (PNRR), Mission 4, Component 2, Investment 1.2, project ``GlassNEXT'', CUP C93C24008240006, and by the STARS@UNIPD program. We acknowledge the ESRF for provision of synchrotron radiation facilities under proposal No.\ SC-5581. SM, DM, LP, FD and GM acknowledge support by the project GLAXES ERC-2021-ADG (grant agreement No.\ 101053167) funded by the European Research Council, European Union. SSS acknowledges the Danish Agency for Science, Technology, and Innovation for travel funding through DanScatt. EP acknowledges financial support from Mar\'ia de Maeztu CEX2023-001300-M and PID2023-146623NB-I00 funded by MICIU/AEI/10.13039/501100011033 and FEDER, UE.

\paragraph{Author contributions}
P.S. and B.B. conceived the study and designed the experiments. P.S. and G.M. designed the combined fast scanning calorimetry and X-ray scattering setup, with assistance from M.D.M. B.B., S.M., D.M., L.P., F.D., S.S.S., M.D.M., and P.S. performed the X-ray scattering measurements. B.B. performed the calorimetric measurements. E.P. provided the samples. P.S. and B.B. developed the analysis procedures and analyzed the scattering and calorimetric data. P.S. and G.M. supervised the project. P.S. and B.B. wrote the initial manuscript, and all other authors reviewed and revised it.
\end{acknowledgments}

\section{Methods}

\subsection{Sample preparation and combined FSC--X-ray scattering measurements}
For all measurements reported in this work, a small piece of sample is cut from a ribbon of  Pd$_{42.5}$Cu$_{30}$Ni$_{7.5}$P$_{20}$ (at.\%) produced by melt-spinning and subsequently placed on a Mettler–Toledo UFH1 chip for calorimetric and X-ray scattering measurements. The size of all measured samples is on the order of a few tens of micrometers in each dimension.

\emph{In-situ} X-ray total scattering measurements combined with FSC are performed using our setup demonstrated at beamline ID15A at the ESRF~\cite{Sun2025ACalorimetry}. The setup enables low-background X-ray total scattering measurements, allowing small structural changes to be accurately resolved even in amorphous materials during and after rapid thermal scans.

In order to prepare the sample in precise thermodynamic states, we follow the thermal protocol illustrated in Figure~\ref{fig:thermal_program}. Here, all temperature ramps reach $T_r = 703$\,K, well above the calorimetric glass transition temperature, $T_g$, reported to range from 566\,K to 580\,K~\cite{Georgarakis2011VariationsVitrification,Haruyama2014VolumeGlass,Qiao2014CharacteristicsLiquids}, and are held there for 0.1\,s so that the sample reaches equilibrium. 
The thermal protocol consists of a sample preparation ramp, shown in red and performed at a rate of $10^4$\,K/s, followed by a probing ramp at $10^3$\,K/s, shown in blue. 
Two sets of these ramps are performed: one without and one with an aging isotherm during the sample preparation ramp. The protocol is repeated for various aging times $t_\mathrm{ag}$, ranging from milliseconds to hours, and aging temperatures $T$, ranging from 423\,K to 613\,K, from well below to well above $T_g$.

X-ray diffraction patterns are acquired during the aging isotherm, marked by the gray shaded region in the figure. 
The calibration of the detector position is performed by collecting diffraction data with polycrystalline CeO$_2$ placed at the sample position.
The 2D diffraction patterns are corrected for the flat field and azimuthally integrated using the \emph{pyFAI} library~\cite{Kieffer2025ApplicationCrystallography}, obtaining the diffracted intensity as a function of the momentum transfer, $Q$. Then, we obtain the structure factor $S(Q)$ using the \emph{PDFgetX3} software~\cite{Juhas2013}, with the following parameters: $Q_\mathrm{min}=0.7\,\mathrm{\r{A}}^{-1}$, $Q_\mathrm{max}=18.0\,\mathrm{\r{A}}^{-1}$, and $\mathrm{rpoly}=0.9$ (default value). Finally, the $S(Q)$ curves as a function of time $t$ at each aging temperature are binned on a logarithmic scale in $t$ for further analysis.

\subsection{Enthalpy measurements}
\begin{figure}
    \centering
    \includegraphics{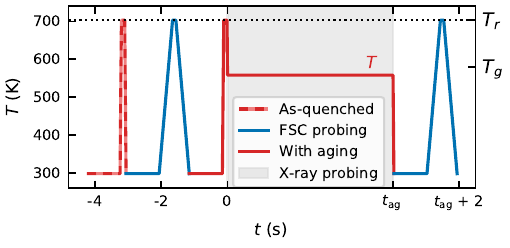}
    \caption{\textbf{Thermal protocol.}
    The two sample-preparation ramps are shown in red, one without and one with the aging isotherm. XRD measurements are acquired only during the aging isotherm (gray-shaded region). In contrast, FSC measurements are performed during the probing ramps (blue), following either the aged or non-aged preparation. The heating and cooling rates are fixed at $10^4$\,K/s during sample preparation (red ramps) and at $10^3$\,K/s for probing (blue ramps).}
    \label{fig:thermal_program}
\end{figure}

The enthalpy difference during physical aging, $\Delta H$, is measured in the following way. First, we take the difference in the heat flow during the heating ramps in the probe (blue curves in Figure~\ref{fig:thermal_program}), for the as-quenched and aged samples. This difference is then converted to a heat capacity, $\Delta c_p(T)$, by dividing by the probing rate and the sample mass. The latter is determined by comparing the corrected heat flow data of the as-quenched sample with literature data for the heat capacity of the glass from Ref.~\cite{Haruyama2007FreeGlasses} measured with conventional differential scanning calorimetry, following the procedure described in Ref.~\cite{Sonaglioni2023}. The mass of sample for the enthalpy measurements reported here is about 579\,ng.
To account, to first order, for a small residual mismatch between the baselines of the as-quenched and aged samples, a linear correction is applied (cf.\ Ref.~\cite{Baglioni_2024}), fixing the difference to be zero in the supercooled liquid and in the room-temperature glass, where the two preparations are expected to result in the same $c_p$. The fitted linear function is then subtracted from the $c_p$ difference. Finally, the corrected signal is integrated over the full temperature range to obtain $\Delta H$.

\subsection{Calculation of changes in PDF}
The changes in the PDF are obtained in the following way: first, because to first order the changes in the intensity profile $I(Q)$ appear to be single-mode, we extract the main component $\Delta I(Q)$ using a Weighted Alternating Least Squares method that takes into account error in the data. The extracted component is then smoothed using a univariate spline function and normalized so that its integral in the range for $S_2$, $4.5\,\mathrm{\AA}^{-1}\leq Q \leq 5.0\,\mathrm{\AA}^{-1}$, is 1\% of that of the baseline $I(Q)$. Because in the regime of small variations, the change in PDF $\Delta G(r)$ is proportional to $\Delta I(Q)$, this normalization helps to quantitatively compare the results at different temperatures. Finally, we use \emph{PDFgetX3}~\cite{Juhas2013} to extract $G(r)$ and its relative change defined as
\[
\Delta G(r) \equiv G(r; I+\Delta I) - G(r; I).
\]

%

\appendix
\setcounter{figure}{0}
\renewcommand{\thefigure}{S\arabic{figure}}

\begin{figure*}
    \centering
    \includegraphics{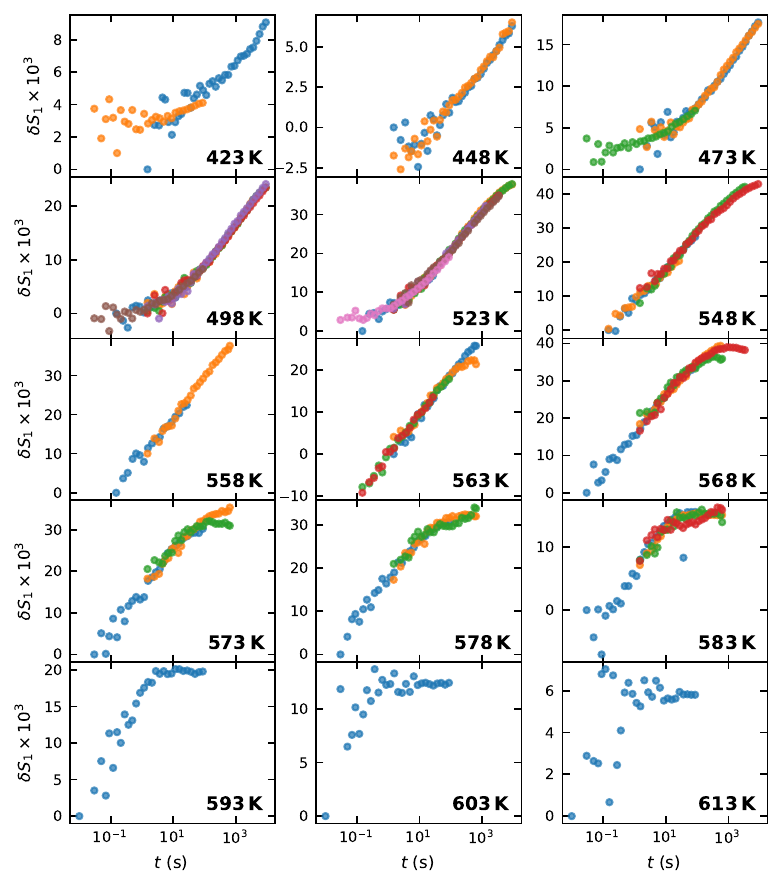}
    \caption{\textbf{Combining $\delta S_1$ datasets.} Each panel shows the $\delta S_1$ values obtained during aging at the indicated temperature. Different colors distinguish independently acquired datasets, corresponding either to a single scan or to the average of a series of repeated scans; averaging is used particularly at short aging times to improve statistics. A constant vertical offset in $\delta S_1$ is applied to each dataset as needed to align measurements in overlapping time ranges. This accounts for sample-to-sample differences and possible errors in the absolute $S_1$ value. The datasets at each temperature are combined and re-binned on a log scale for further analysis.}
    \label{fig:S1_combined}
\end{figure*}

\begin{figure*}
    \centering
    \includegraphics{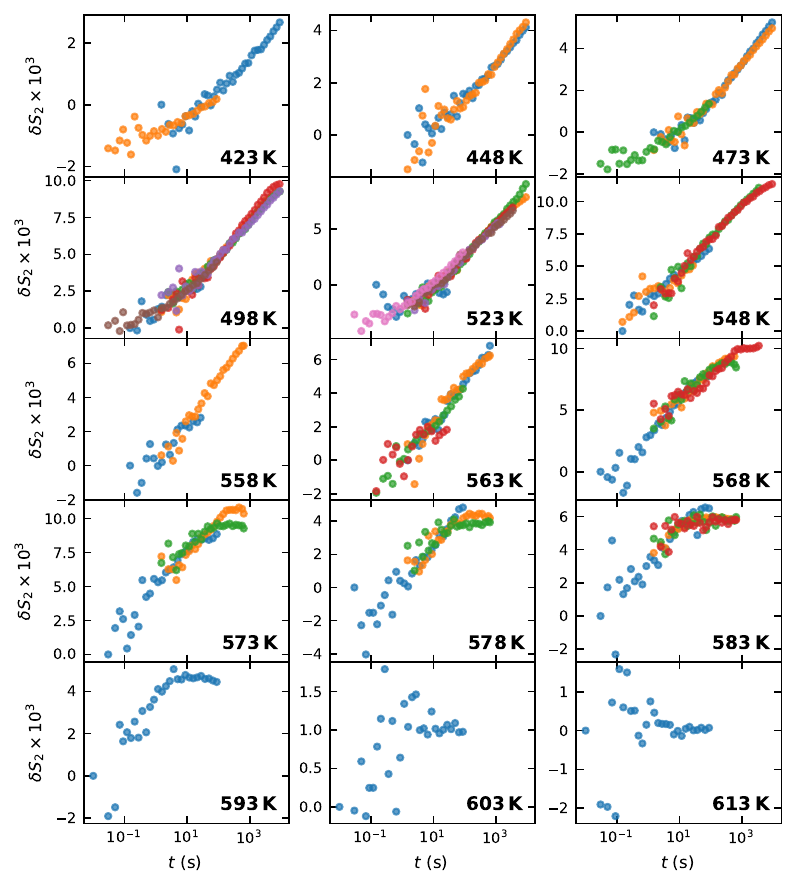}
    \caption{\textbf{Combining $\delta S_2$ datasets.} Same as Figure~\ref{fig:S1_combined}, but for $\delta S_2$.}
    \label{fig:S2_combined}
\end{figure*}

\begin{figure*}
    \centering
    \includegraphics{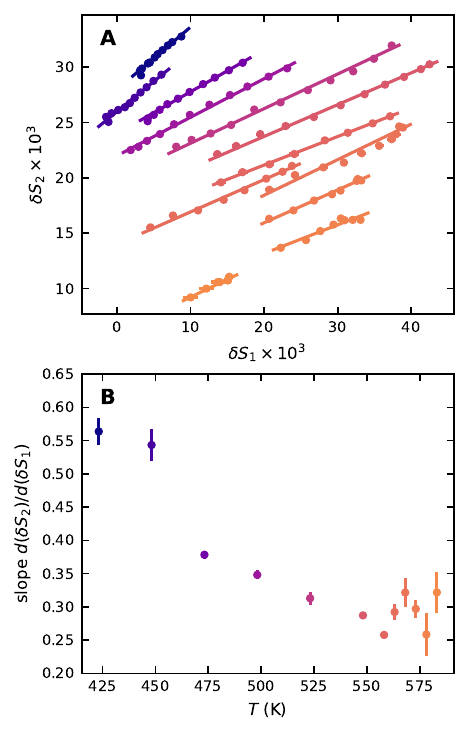}
    \caption{\textbf{Relation between $\delta S_1$ and $\delta S_2$.} (A) $\delta S_1$ plotted against $\delta S_2$ at different aging temperatures indicated by different colors (same as in panel B), together with linear fits. To reduce the influence of noise at short aging times, only data for $t >3$\,s are included. Vertical offsets are applied between temperatures for clarify. (B) The slope $d(\delta S_2)/d(\delta S_1)$ obtained from the linear fits in panel A, plotted as a function of temperature. Note that the linear fits are only intended as a first-order description of the relation between $\delta S_1$ and $\delta S_2$.}
    \label{fig:slope_dS2dS1}
\end{figure*}

\end{document}